\documentclass[showpacs,preprintnumbers,amsmath,amssymb,floatfix]{revtex4}
\usepackage[applemac]{inputenc}
\usepackage[T1]{fontenc} 
\usepackage{subfigure}
\usepackage{amsmath}
\usepackage{amsfonts}
\usepackage{amssymb}
\usepackage[dvips]{graphicx}
\usepackage[scr=rsfs]{mathalpha}
\usepackage{relsize}
\usepackage{mathtools,slashed}
\usepackage{ulem}
\usepackage[colorlinks]{hyperref}
\usepackage[usenames,dvipsnames]{color}
\usepackage{bbm}
\def\beq{\begin{equation}}
\def\eeq{\end{equation}}
\def\bea{\begin{eqnarray}}
\def\eea{\end{eqnarray}}
\def\ba{\begin{array}}
\def\ea{\end{array}}
\def\nn{\nonumber}
\def\ame{&=&}

\def\part{\partial}

\def\Tr{\mbox{Tr}}

\def\ie{{\it{i.e. }}}
\def\nn{\nonumber}

\def\R{\ensuremath{\mathbb{R}}}

\def\H{\ensuremath{\mathbb{H}}}

\newcommand{\bS}{{\mathbb S}}
\newcommand{\bI}{{\mathbb I}}

\newcommand{\cD}{{\mathcal{D}}}
\newcommand{\cE}{{\mathcal{E}}}

\newcommand{\cM}{{\mathcal{M}}}

\newcommand{\cR}{{\mathcal{R}}}

  \def\d{\mathrm{d}}

\newcommand{\SO}{{\rm SO}}    %special orthogonal group
\newcommand{\SU}{{\rm SU}}    %special unitary group
\newcommand{\so}{{\mathfrak{so}}}  %special orthogonal Lie algebra
\newcommand{\su}{{\mathfrak{su}}}  %special unitary Lie algebra
\def\b{\beta}

\def\b1{\mathbb{1}} % identity
\def\R{\mathbb{R}} % Reals
\def\bS{\mathbb{S}}

\def\cD{\mathcal{D}}
\def\cE{\mathcal{E}}

\def\cM{\mathcal{M}}

\def\cR{\mathcal{R}}

\def\cY{\mathcal{Y}}

\def\so{\mathfrak{so}}

\def\su{\mathfrak{su}}

\begin{document}

\preprint{UdeM-GPP-TH-25-304}

\title{Physics on manifolds with exotic differential structures}
\author{Ulrich Chiapi-Ngamako$^{1}$}
\email{ulrich.chiapi.ngamako@umontreal.ca}
\author{M. B. Paranjape$^{1,2,3}$}
\email{paranj@lps.umontreal.ca}

\affiliation{$^1$Groupe de physique des particules, Département de
physique,  Université de Montréal, Campus MIL, 1375 Av Thérèse-Lavoie-Roux,, Montr\'eal, Qu\'ebec, Canada, H3B 2V4 }
\affiliation{$^2$  Centre de recherche mathématiques and Institut Courtois, Universit\'e de Montr\'eal.}
\affiliation{$^3$ Department of Physics, University of Auckland, Auckland, New Zealand, 1010}

\begin{abstract}
A given topological manifold can sometimes be endowed with inequivalent differential structures.  Physically this means that what is meant by a differentiable function (smooth) is simply different for observers using inequivalent differential structures.  {The 7-sphere, $\bS^7$, was the first topological manifold where the possibility of inequivalent differential structures was discovered \cite{Milnor}.} In this paper, we examine the import of inequivalent differential structures on the physics of fields obeying the  Dirac equation on $\bS^7$.  { $\bS^7$ is a fibre bundle of the 3-sphere as a fibre on the 4-sphere as a base.  We consider the Kaluza-Klein limit of such a fibre bundle which reduces to a SO(4) Yang-Mills gauge theory over  $\bS^4$.  We find, for certain specific symmetric set of gauge potentials, that the spectrum of the Dirac operator can be computed explicitly, for each choice of the differential structure.  Hence  identical topological manifolds have different physical laws.  We find this the most important conclusion of our analysis.}
\end{abstract}
\maketitle
\section{Introduction}
Almost all of physics relies on being able to take the derivative of some relevant real-valued function.  For a general, $n$ dimensional manifold, the notion of what is a differentiable function  on the manifold depends on the set of charts (continous, invertible maps (homeomorphisms) from open sets in $\R^n$ to the manifold) that cover (each point in the manifold is the image of a point in some chart) the manifold.  A function that is defined on the manifold, is pulled-back (\ie defined by the composition of the map corresponding to the chart with the function on the manifold) to a local set in $\R^n$, and the derivative is accordingly defined by the derivative of the pulled-back function on the local  set in $\R^n$.  

However, a manifold is only completely defined by the union of open charts that cover the topological space.  Where two charts intersect, we can define a function from $\R^n\to\R^n$, the so-called transition functions, using one chart to go to a point in the intersection on the manifold and then using the inverse of the second chart to return to $\R^n$.  One can impose conditions on 
these transition functions.  A topological manifold requires only that the transition functions be continous.  A smooth manifold requires that the transitions functions be infinitely differentiable with infinitely differentiable inverse.  An atlas of a $C^N$ manifold consists of the union of all charts such that the transition functions and their inverses are $C^N$, \ie $N$ times continuously differentiable.  We say that the manifold admits a $C^N$ differentiable structure.  It is then clear that a $C^0$ manifold, \ie simply a topological manifold, admits a much larger atlas than a $C^\infty$ manifold, the transition functions need only be continous.  Indeed then, it is not impossible to imagine that inequivalent subsets of the charts of a topological manifold could give rise to inequivalent $C^\infty$ structures, \ie give rise to different, $C^\infty$ atlases that cannot be combined while maintaining the $C^\infty$ of each other.  

Milnor\cite{Milnor} gave the first example of such a case for $\bS^7$.  Subsequently Milnor and Kervaire \cite{Milnor-Kervaire} analyzed the possibility of inequivalent differentiable structures on all possible finite dimensional manifolds.  These examples were mathematical oddities and did not seem very relevant to physics.  However, in the 80s,  Freedman's analysis \cite{Freedman1,Freedman} and Donaldson's subsequent analysis \cite{Donaldson} of the moduli spaces of instantons on $\R^4$  made the shocking discovery that $\R^4$ admits inequivalent differentiable structures, and that $\R^4$ is very special in that respect, all other $\R^N$s admit only one differentiable structure.  This prompted an intriguing speculation by Taubes \cite{Taubes} about how physical systems choose the differentiable structure and what would be import of the inequivalent differentaible structures on the physics.   We make some inroads into answering this sort of question by studying physics on the original, exotic $\bS^7$s of Milnor.   Although there has been some work done on physics on exotic $\bS^7$s, and exotic manifolds in general, see these references for a partial list \cite{7s1,7s2,7s3,7s4,7s5,7s6,7s7,7s8} and the references within, we find that the nature of these mathematical oddities is not generally understood in the theoretical physics community.  A very recent article that mirrors our analysis closely, especially concerning the Kaluza-Klein approach, is available here \cite{7s9}.

\section{The Exotic $\bS^7\rm\bf s$ of Milnor}
\subsection{Manifolds homeomorphic to $\bS^7$}
The standard, unit  $\bS^7$ is defined by the set of points in $\R^8$ with Cartesian coordinates $(x_1,x_2,\cdots,x_8)$ such that 
\beq
x_1^2 + x_2^2 +\cdots + x_8^2=1
\eeq
and the differential structure is that induced by the unique, differential structure of $\R^8$.   To obtain the exotic $\bS^7$s, Milnor used the generalizations of the Hopf fibering that gives $\bS^7$ as an $\bS^3$ bundle over $\bS^4$.  

The standard Hopf fibering of $\bS^7$ corresponds to using two fundamental charts to describe the manifold.
We use the coordinates
\beq
(u,v)\in \bS^7 \ni u\in \bS^4, \quad v\in \bS^3.
\eeq
Then it is convenient to use the quaternions, $u\in \H$ where $\H$ corresponds to the set 
\bea
&u=u_0+iu_1+ju_2+ku_3, \quad  u_0, \cdots u_3\in\R \nn\\ 
&i^2=j^2=k^2=-1\nn\\
&ij=k,jk=1, ki=j\nn\\
&ij=-ji, jk=-kj, ki=-ik
\eea
The quaternions form a non-commutative field, $|u|= \sqrt{\Sigma_iu_i^2}$ and with the definition $\bar u=u_0-iu_1-ju_2-ku_3$ the inverse is given by $\frac{1}{u}=\frac{\bar u}{|u|^2}$.

Topologically $\H$ corresponds to $\R^4$, hence the quaternionic coordinates can be thought of as the coordinates coming from stereographic projection of $\bS^4$ onto $\R^4=\H$.  $v$ the coordinate on $\bS^3$ can be identified with the set of unit quaternions, $v=v_0+iv_1+jv_2+kv_3$ with $v_0,\cdots v_3$ restricted to a three ball of unit radius and $v_0=\pm\sqrt{1-(v_1^2+v_2^2+v_3^2)}$.  The fundamental set of charts are given by 
\beq
(u,v) \quad {\rm and}\quad (u',v')\label{5}
\eeq
where the coordinates $u$ correspond to stereographic projection from the north pole of $\bS^4$ along with the cartesian product of the coordinates $v$ on $\bS^3$ while the  coordinates $u'$ correspond to stereographic projection from the south pole of $\bS^4$, again with a cartesian product with coordinates $v'$ on $\bS^3$.  The transition functions, corresponding to the (generalized) Hopf fibration, are then defined in terms of quaternions, 
\beq
(u',v')=\left(\frac{u}{|u|^2},\frac{u^hvu^l}{|u|^{h+l}}\right)\label{55}
\eeq
or inversely
\beq
(u,v)=\left(\frac{u'}{|u'|^2},|u'|^{(h+l)}(u')^{-h}v'(u')^{-l}\right) .\label{66}
\eeq
This standard Hopf fibration corresponds to $h=1$, $l=0$ and gives rise to $\bS^7$ analogously to the standard Hopf fibration of $S^1$ on $\bS^2$ giving rise to the 3-sphere.  However, for other values of $h$ and $l$,  generalized fibre bundles with transition functions defined by Eqn. \eqref{55} and Eqn. \eqref{66} give rise to new 7-dimensional manifolds.  Note that arbitrary powers, including inverse powers, of quaternions make perfect sense, $h$ or $l$ can be positive or negative.  

Amazingly, for the case $h+l=1$, the manifolds are topologically homeomorphic to the standard $\bS^7$.    For this case, the transition functions become
\bea
(u',v')\ame\left(\frac{u}{|u|^2},\frac{u^hvu^l}{|u|}\right)\nonumber\\
(u,v)\ame\left(\frac{u'}{|u'|^2},{|u'|}(u')^{-h}v'(u')^{-l}\right).\label{1}
\eea
To prove this, Milnor \cite{Milnor} invoked Morse theory \cite{Morse} and specifically Reeb's theorem \cite{Reeb} which states if a function can be defined on a $d$-dimensional, compact manifold which has exactly two, non-degenerate critical points, then the manifold is homeomorphic to a d-dimensional sphere.  Morse theory relates the critical points of a function to the minima, maxima and topological handles (minimaxes) on the manifold.  For a compact manifold with exactly two critical points, these critical points have to be the global minimum and the global maximum,  there can be no handles.  Reeb's theorem then states that the manifold has to be topologically a sphere.  For the case $h+l=1$,  Milnor \cite{Milnor} exhibited the following Morse function
\beq
f(u,v)= \frac{{\cal R} (v)}{\sqrt{1+|u|^2}}
\eeq
where ${\cal R} (v)$ stands for the real part of $v$, and showed that it has exactly two critical points.  ${\cal R} (v)=v_0=\pm{\sqrt{1- (v_1^2+v_2^2+v_3^2)}}$.  We can see this by calculating the derivatives of the Morse function in the coordinate system given by the $u_i$ with $i=1,2,3,4$ and $v_i$ with $i=1,2,3$.  For a critical point we need
\bea
\frac{\partial}{\partial u_i}f(u,v)= \frac{-{\cal R} (v)}{(1+|u|^2)^{3/2}}u_i=0\nn\\
\frac{\partial}{\partial v_i}f(u,v)= \frac{-v_i}{{\cal R}(v){\sqrt{1+|u|^2} }}=0
\eea
which means $u_i=0$ and $v_i=0$, which implies $(u,v)=(0,\pm 1)$.  These are the only two critical points in the northern patch.  For the southern patch, we have
\bea
f(u',v')\ame\frac{{\cal R} (|u'|{u'}^{-h}v'{u'}^{-l})}{\sqrt{1+1/|u'|^2}}\nn\\
\ame\frac{{\cal R} ({|u'|}{u'}^{-(h+l)}v')}{\sqrt{1+1/|u'|^2}}\nn\\
\ame\frac{|u'|{\cal R} ({u'}^{-1}v')}{\sqrt{1+1/|u'|^2}}\nn\\
\eea
where we have used $h+l=1$ and that ${\cal R} $ is cyclic.  Then using ${\cal R}(q^{-1})={\cal R} (\bar q/|q|^2)={\cal R} (q/|q|^2)$ for any quaternion $q$ and $|v'^{-1}u'|=|u'|$ as $v'$ is a unit quaternion, we have
\bea
f(u',v')\ame\frac{|u'|{\cal R} (v'^{-1}u')}{|u'|^2\sqrt{1+1/|u'|^2}}\nn\\
\ame\frac{{\cal R} (u'v'^{-1})}{\sqrt{1+|u'v'^{-1}|^2}}.
\eea
Now $u'v'^{-1}$ is a perfectly general, independent quaternion, call it $u''=u''_0+iu''_1+ju''_2+ku''_3$.  
Then
\beq
f(u',v')=\frac{{\cal R}(u'')}{\sqrt{1+|u''|^2}}=\frac{u''_0}{\sqrt{1+{u''_0}^2+{u''_1}^2+{u''_2}^2+{u''_3}^2}}
\eeq
It is easy to see that the derivative of this function with respect to $u''_0$ never vanishes
\bea
\frac{\partial f(u',v')}{\partial u''_0}\ame \frac{1}{\sqrt{1+{u''_0}^2+{u''_1}^2+{u''_2}^2+{u''_3}^2}}-\nn\\
&-&\frac{{u''_0}^2}{(1+{u''_0}^2+{u''_1}^2+{u''_2}^2+{u''_3}^2)^{3/2}}\nn\\
\ame \frac{(1+{u''_1}^2+{u''_2}^2+{u''_3}^2)}{(1+{u''_0}^2+{u''_1}^2+{u''_2}^2+{u''_3}^2)^{3/2}}>0.\nn\\
\eea
Hence the function has no critical points in the southern patch and exactly two critical points in the northern patch, \ie two critical points that are easily seen to be non-degenerate.  Hence by Morse theory and specifically Reeb's theorem, the manifold is homeomorphic to the standard $\bS^7$.  Let us call the manifolds $M^7_k$ where $h+l=1$ but $h-l=k$.  
\subsection{Existence of diffeomorphically inequivalent $\bS^7$s}
Then the proof that some of these fibre bundles are not diffeomorphic to the standard $\bS^7$ follows from the Hirzebruch signature theorem \cite{Hirzebruch}.  One assumes that $M^7_k$ are indeed diffeomorphic to the standard $\bS^7$ and then we obtain a contradiction. 

An integer valued, modulo 7, topological invariant, $\lambda(M^7_k)$, can be defined for the manifolds $M^7_k$.  First we construct a smooth, 8-dimensional manifold, $B^8$, whose boundary is given by $M^7_k$.  $B^8$  always exists by a theorem of Thom \cite{Thom} given $M^7_k$ is closed, oriented and with vanishing 3rd and 4th cohomology groups.    That these cohomology groups vanish is clear because $M^7_k$ is homeomorphic to $\bS^7$, and $\bS^7$ only has non-vanishing cohomology classes $H^0(\bS^7)$ and $H^7(\bS^7)$.  The standard $\bS^7$ is the boundary of the standard 8-disc $D^8$.  As $M^7_k$ is homeomorphic to the standard $\bS^7$, {\it and now we assume diffeomorphic},  we can smoothly glue together $D^8$ to $B^8$ on their boundary to form a smooth, closed 8-dimensional manifold which we will call $W^8_k$.  Then the Hirzebruch signature theorem says
\beq
\sigma(W^8_k)=\frac{1}{45}(7p_2(W^8_k) -p_1^2(W^8_k))
\eeq
where $p_1$ and $p_2$ are the first and second Pontrjagin class respectively.  The signature $\sigma(W^8_k)=\pm 1$, choose +1, then we have
\beq
45 +p_1^2(W^8_k)=0 \,\,\,{\rm modulo }\,\,\, 7.
\eeq
Then it is incumbent on us to compute only $p_1^2(W^8_k)$, which is found by Milnor, \cite{Milnor}, to be $4k^2$.  Thus we get the equation
\beq
45 +4k^2=0 \,\,\,{\rm modulo }\,\,\, 7 \quad\ie\quad 3 +4k^2=0 \,\,\,{\rm modulo }\,\,\, 7.
\eeq
$k=\pm 1$ obviously is a solution, but $k=2,3,4,5$ are easily seen not to satisfy this equation, which is a contradiction.  \beq
2\,\, {\rm and}\,\, 5=\pm 2\,\,\,{\rm modulo }\,\,\, 7\Rightarrow  3+4\cdot 4=19=5\ne 0 \,\,\,{\rm modulo }\,\,\, 7
\eeq 
and
\beq
3\,\, {\rm and}\,\, 4=\pm 3\,\,\,{\rm modulo }\,\,\, 7\Rightarrow 3+4\cdot 9=39 =4\ne 0 \,\,\,{\rm modulo }\,\,\, 7 .
\eeq 
Therefore, the assumption that we made, that $M^7_k$ is diffeomorphic to the standard $\bS^7$ has to be false for the cases $k=2,3,4,5\,\, {\rm modulo}\,\, 7$ and as such there exist exotic $\bS^7$s that are homeomorphic to the standard $\bS^7$, topologically the same, but that cannot be diffeomorphic to the standard $\bS^7$.

This result is rather astonishing.  Two manifolds which have the same notion of continuous functions do not have the same notion of differentiable functions.  The fundamental question arises, what part of physical reality depends only on the notion of continuity, and not on the notion of differentiabilty.   All kinds of physical phenomena do not depend on the global differential structure of the manifold on which the phenomena occurs.  The diffeomorphically inequivalent $\bS^7$s all admit smooth metrics, which give a notion of length scale.  All phenomena which occurs esentially locally, such as crystal growth or any biological phenomena for example, are simply identical in any spacetime that is smooth, but where the length scale of the physical phenomena is small compared to the length scale over which the differential structures varies.   Our diffeomorphically inequivalent $\bS^7$s are of course locally flat when equipped with a metric, and the inequivalent differential structures occur only because of global obstructions.  Hence, physical phenomena which occur over length scales small compared to the length scale of the variation of the differential structure are bound to be identical.  

However, all of classical or quantum mechanics depends on the notion of differentiability.  Hence there will clearly be criteria by which one could physically discern between topologically equivalent manifolds which are not diffeomorphic.  This is what we endeavour to find in the rest of this paper.  We will look at the spectrum of the Dirac operator on the different, exotic $\bS^7$s compared with the operator defined on the standard $\bS^7$.  The spectrum of the operator, especially for the low-lying modes will clearly be of physical importance and will give a tangible criterion with which to discern between exotic and standard $\bS^7$s.  The metric on $\bS^7$s can be chosen to correspond to a Kaluza-Klein reduction.  This does not affect the global topology nor the differential structure.  In this reduction, the metric on the $\bS^3$ is taken so that the size of the 3-sphere is very small compared to the size of the base, $\bS^4$.  Then the effective theory we are left with is a Einstein-Yang-Mills theory on the $\bS^4$ base.  Such a theory could be quite relevant to our 4-dimensional physical world.  

\section{Kaluza-Klein reduction}
{Having established that the manifolds $M^7_k$, for $h+l=1$ are all homeomorphic to $\bS^7$, we will imagine the Kaluza-Klein reduction of the manifolds, \cite{Straumann,SalamStrathdee,WittenKK}.  {It is important to underline that} such a reduction maintains the topology and the differential structure of the manifold{, it remains topologically a $\bS^7$}.   {This is specifically because the argument using Morse theory and Reeb's theorem given above to demonstrate that the manifolds are all homeomorphic to $\bS^7$ does not make any reference to a metric, the proof is simply based on the existence of a differentiable function with strictly non-degenerate critical points (a Morse function).  The topological manifold $\bS^7$, and evidently $M^7_k$ as these are homeomorphic, is well known to be a spin manifold.  For this to be the case the first and second Stiefel-Whitney classes of the tangent bundle of $M^7_k$,  must vanish, \ie $H^1(M^7_k, \mathbb Z_2)=H^2(M^7_k, \mathbb Z_2)=0$.  These Stiefel-Whitney classes are contained in the corresponding integral cohomology classes.  The only non-vanishing integral cohomology classes of a topological $n$-sphere are $H^0(\bS^n, \mathbb Z)=\mathbb Z=H^n(\bS^n, \mathbb Z)=\mathbb Z$ hence specifically, $H^1(M^7_k, \mathbb Z)=H^2(M^7_k, \mathbb Z)=0$.  Therefore the corresponding Stiefel-Whitney classes must also vanish.}   }

{There is nothing mysterious about the manifolds $M^7_k$.  They are simply defined as $\bS^3$ bundles over $\bS^4$ with the topological twisting specified by the two integers $h$ and $l$.  They inherit the natural differential structure defined by the atlas containing the charts in their definition, \cite{Milnor}.  The only surprising thing about these manifolds is the Milnor observation that they can be homeomorphic but not diffeomorphic to the standard $\bS^7$. The fibre, $\bS^3\equiv SU(2)$, is part of the manifold and there is no obstruction to putting the spherically symmetric Cartan-Killing metric on it.  Overall, the metric must be smooth on the full manifold with the differential structure defined by the transition functions (and the corresponding atlas).}  

The metric on the corresponding fibre bundle can be taken to be \cite{Straumann,SalamStrathdee,WittenKK}:
\beq
g= g_{\bS^4}+k_{ij}(\theta^i-K^i_a A^a_\mu dy^\mu)\otimes (\theta^j-K^j_b A^b_\mu dy^\mu)\label{3}
\eeq
where $g_{\bS^4}$ is the metric on the base $\bS^4$ with coordinates $y^\mu$, $k_{ij}$ are the components of the Cartan-Killing metric on the fibre $\bS^3$, $K^i_a$ are the components of the Killing vectors that describe the isometries of the fibre, $\SO(4)$, $\theta^i$ are the components of the dreibein (triad) one forms on the fibre {which can be taken simply as the Maurer-Cartan one-forms of the group manifold} and the fieldsm $A^a$, are part of the full metric.   However, in the Kaluza-Klein reduction, they become  the components of a Yang-Mills gauge field corresponding to the gauge group given by the isometries of the fibre, $\SO(4)$, {
 $a$ takes values $1,\cdots , 6$ while $i$ takes the values  $1,2,3$.  

Explictly,with the standard Hopf  coordinates  
\bea
x^1+ix^2=\cos2\theta e^{i(\psi+\phi)/2}\\
x^3+ix^4=\sin2\theta e^{i(\psi-\phi)/2}
\eea
the Maurer-Cartan 1-forms are given by
\bea
\theta^1\ame\cos\psi d\theta +\sin\psi\sin\theta d \phi\\
\theta^2\ame -\sin\psi d\theta +\cos\psi\sin\theta d \phi\\
\theta^3\ame \d\psi+\cos\theta d\phi
\eea
and then  $k_{ij}=\delta_{ij}$.  The 6 Killng vectors are given by:
\bea
K_1^idx_i\ame\cos\psi d\theta+\sin\psi\sin\theta d\phi\\
K_2îdx_i\ame-\sin\psi d\theta+\cos\psi\sin\theta d\phi\\
K_3^idx_i\ame d\psi+\cos\theta d\phi
\eea
and 
\bea
K_4^idx_i\ame\cos\phi d\theta-\sin\phi\sin\theta d\psi\\
K_5îdx_i\ame -\sin\phi d\theta-\cos\phi\sin\theta d\psi\\
K_6^idx_i\ame d\phi+\cos\theta d\psi
\eea

The gauge fields come from the metric of Eqn.\eqref{3}.  They are functions only of the coordinates of the base manifold $\bS^4$ and do not depend on the coordinates of the fibre, $\bS^3$.  {A coordinate tranformation on $\bS^4$ that preserves the fibre metrically however transforms it by an isometry exhibits entirely as an $\SO(4)$ gauge transformation on the field $A^a$, see  \cite{Straumann,SalamStrathdee,WittenKK} for the details.}  The gauge field is necessarily present as the manifold is a non-trivial fibre bundle of $\bS^3$ over $\bS^4$.  If the gauge field were absent, the manifold would simply be only the Cartesian product of $\bS^3$ with $\bS^4$, which is not even a standard $\bS^7$.  {The metric \eqref{3} is perfectly valid before any assumption on the size of the $\bS^3$ compared to the size of the $\bS^4$, \ie it is generally valid before any Kaluza-Klein reduction is considered.}

{The Kaluza-Klein reduction means that we choose the metric such that the fibre should be a simple, spherically symmetric $\bS^3$ of {negligible radius compared to the radius of the base also chosen to be simple, spherically symmetric $\bS^4$} and only the isometries of $\bS^3$ and not its deformations can have any impact on the low energy dynamics taking place in the ambient space given by the base, $\bS^4$.  It is not consistent in this limit to think of deformations of the fibre, these would correspond only to very high energy excitations.  Then the only degrees of freedom left from the fibre arise from the liberty to rotate it arbitrarily by the group transformations that are symmetries (isometries) of the metric on the fibre, as we move along the base manifold.  This gives rise to a gauge degree of freedom, the gauge group being in this case $\SO(4)$.  {We want to stress that the effective gauge theory on $\bS^4$ is really just gravity on $\bS^7$.  The gauge fields come from the metric \eqref{3}, and are not arbitrarily added.} }

{The low energy dynamics coming from an assumed Einsteinian dynamics on the original 7 dimensional manifold then simply reduces to 4 dimensional Einstein gravity on $\bS^4$ coupled to $\SO(4)$ gauge fields with Yang-Mills dynamics.  Coordinate transformations transform the metric in the standard fashion, but the subset of coordinate transformations which simply rotate the fibre by an isometry as a function of of the coordinates on the base, simply give rise to a (non-abelian) gauge transformations of the gauge fields.  However, most importantly, due to the exotic differential structure, these gauge fields have to be connections on topologically non-trivial fibre bundles that are distinct from the standard Hopf fibring that gives rise to the standard $\bS^7$.  This means that they must have topological invariants that are distinct from those of gauge fields that would be defined on the standard $\bS^7$ also in the Kaluza-Klein limit. }

The metric on $\bS^4$, $g_{\bS^4}$, can be arbitrary, the simplest to take is the constant curvature metric.  The metric on $\bS^3$, which would be $k_{ij}\theta^i\otimes \theta^j$ if the gauge field $A^a$ were absent, is the Cartan-Killing metric, and it is also of constant curvature.    To make a $\bS^7$, the $\bS^3$ fibre has to be twisted as it goes around the equator of $\bS^4$.  It is the gauge fields that capture the topologically non-trivial structure inherent in the normal and exotic $\bS^7$s, and as such impose global constraints on the possible gauge fields.  In the Kaluza-Klein reduction of the manifold, the base manifold is topologically and differentiably $\bS^4$, but it is locally a direct product with a tiny $\bS^3$ associated with each point of the $\bS^4$.  This $\bS^3$ twists as it is defined over the $\bS^4$.  These twistings, are defined by the generalized Hopf fibrings defined by Eq.\eqref{1},  for $h+l=1$.

The metric can be defined in terms of the vielbeins $e^A_\mu$, $g_{\mu\nu}=\eta_{AB}e^A_\mu \otimes e^B_\nu$, then the spin connection is defined by the equation $de^A+\Omega^A_{\,\, B}\wedge e^B=0$ and the curvature 2-form is defined by $R^A_{\,\, B}=d\Omega^A_{\,\, B}+\Omega^A_{\,\, C}\wedge \Omega^C_{\,\, B}=\frac{1}{2}R^A_{\,\, BCD}e^C\wedge e^C$ where all indices $\mu$ and $A$ go from 1 to 10{, seen as a metric on the 10 dimensional space corresponding to the $\SO(4)$ principal bundle over $\bS^4$, with a 6 dimensional fibre and a 4 dimensional base space, giving a total of 10 dimensions}.  It is well understood \cite{Straumann}, that with the metric of the form Eqn.\eqref{3}, the scalar curvature is simply given by
\beq
R=R_{\bS^4}+R_{\bS^3}+{\cal L}_{YM}
\eeq
where $R_{\bS^4}$ is the scalar curvature of $g_{\bS^4}$ on $\bS^4$, $R_{\bS^3}$ is the scalar curvature of $k_{ab}$ on $\bS^3$ and ${\cal L}_{YM}$ is the Yang-Mills Lagrangian for the gauge field $A^a$ on $\bS^4$.  

The gauge field must be consistent with the bundle structure defined by $h$ and $l$.  This means that the transition functions for the gauge fields between the northern patch and the southern patch must reflect the values of $h$ and $l$, specifically, the action on the fibre from Eqn.\eqref{1} is given by
\beq
v'=\hat u^h v \hat u^l \label{4}.
\eeq
{With that constraint, the gauge fields are consistent with the transition functions that are the fundamental reason why the different manifolds $M^7_k$ have inequivalent differential structures from the standard $\bS^7$, and correspondingly, the gauge fields and the metric are consistent with the differential structure. } The bundle is an $\bS^3$ bundle over $\bS^4$, the isometry group of $\bS^3$ being $\SO(4)$, therefore we actually construct an $\SO(4)$ principal bundle over $\bS^4$.  The defining representation consists of $4\times 4$ dimensional matrices acting on four dimensional vector in $\R^4$.  The general quaternionic transformation
\beq
x'=\hat q x \hat r ,
\eeq
with $x=x_0+ix_1+jx_2+kx_3$, $\hat q = \cos\theta+\sin\theta \hat\theta\cdot\vec i$ and $\hat r = \cos\zeta+\sin\zeta \hat\zeta\cdot\vec i$ where  $\vec i \equiv (i,j,k)$ of the vector of the fundmental quaternions, can be written as
\beq
{x'}^\mu=(R_L R_R^T)^\mu_{\,\,\,\nu} x^\nu
\eeq
where $R_R^T$ is the transpose (hence the inverse) of the orthogonal matrix $R_R$ and $\mu \, ,\,\nu\in \,1,2,3,4$.  $R_L$ and $R_R$ are respectively the left and right isoclinic decompositions of the fundamental representation of $\SO(4)$.  Here we can take the explicit representations, 
\beq
R_L=\cos\theta+\sin\theta \hat\theta\cdot\vec T_L\quad R_R=\cos\zeta+\sin\zeta \hat\zeta\cdot\vec T_R
\eeq
with $\theta\,\zeta\in (0,\pi)$ to fully cover the unit quaternions, and the (anti-hermitian) generators
\beq
 T^1_L=-i \bI\otimes\tau^2\quad T_L^2=-i\tau^2\otimes\tau^3\quad T_L^3=-i\tau^2\otimes\tau^1
\eeq
and
\beq
 T^1_R=i \tau^3\otimes\tau^2\quad T_R^2=i\tau^2\otimes\bI\quad T_R^3=i\tau^1\otimes\tau^2\label{rightisoclinic}
\eeq
where $\tau^i$ are the Pauli matrices.  The generators $T^i_L$ and $T^i_R$ mutually commute and each provide a $4\times 4$ representation of the fundamental quaternions.  Furthermore, $T^i_L/2$ and $T^i_R/2$ are the generators of two independent, reducible representations of $\SU(2)$, the representation $\frac{1}{2}\oplus\frac{1}{2}$.  

For our purposes, from Eqn.\eqref{4}, we have $\hat q\to\hat u^h=\cos (h\theta)+\sin (h\theta) \hat\theta\cdot\vec i$ while $\hat r\to \hat u^l=\cos (l\theta)+\sin (l\theta) \hat\theta\cdot\vec i$.  Then with $R=R_LR^T_R$ we can take the gauge field to be zero in the northern patch, and which satisfies at the equator $A'$ of the southern patch defined as
\beq
A' =R^T (A +d) R
\eeq
and $A'$ is simply switched of to zero as we go the the south pole.  Such a gauge field will not be a solution of the Yang-Mills equations, not have any particular symmetry property, however, it will be consistent with the topological constraints imposed by the bundle structure.  Indeed, the topological number $h-l$ then shows up through the topological invariant called the Pontrjagin number of the gauge field (which is anti-hermitean), $p(A)$:
\bea
p(A)\ame\frac{-1}{16\pi^2}\int_{\bS^4}\epsilon^{\mu\nu\sigma\tau}Tr\left(F_{\mu\nu}F_{\sigma\tau}\right)\nn\\
\ame\frac{-1}{16\pi^2}\int_{\partial S_4=\bS^3}d\sigma_\mu\epsilon^{\mu\nu\sigma\tau}Tr\left( A'_\nu\partial_\sigma A'_\tau +\frac{2}{3}A'_\nu A'_\sigma A'_\tau\right)\nn\\
\ame \frac{1}{24\pi^2}\int_{\partial S_4=\bS^3}d^3x \epsilon^{ijk}Tr\left((R^T\partial_i R)(R^T\partial_j R)(R^T\partial_k R)\right)=2(h-l)
\eea
The factor of two occurs simply because we have a direct sum of two fundamental spin $\frac{1}{2}$ representations in both the left handed and the right handed sectors.  The integral projects to an integral only over the equatorial 3-sphere, which is just the winding number of the map defined by $R$, the left handed part giving $2h$ and the right handed part giving $-2l$.  .  

With the Kaluza-Klein reduction of the exotic $\bS^7$s, we are able to analyze the spectrum of the Dirac operator for $\bS^4$ symmetric gauge fields which are of course consistent with the bundle structure, which we do in the next section.

\section{Spherically Symmetric Instantons and the Dirac Spectrum}
Any gauge field, as long as it satisfies the constraint coming from the global topology, is consistent with the bundle structure, as the example we have chosen above.  {We cannot solve the Dirac equation for any arbitrary gauge field consistent with the topological constraints.  We can however, obtain the exact spectrum of the Dirac operator (squared) if the gauge fields are spherically symmetric and satisfy the Yang-Mills equations of motion.  This is obviously a special case, however, it is an example where it will be clear that the exotic differential structure affects the physically observable aspects of the system.}  There was much work done on ``spherically'' symmetric gauge fields which in fact automatically solve the Yang-Mills equations of motion of the gauge field, and hence are nominally spherically symmetric instantons (\ie exact solutions of the (Euclidean) Yang-Mills equations).  Such gauge field configurations are useful since it is well understood how to find the eigenvalues of the Dirac operator in their presence.  It is these eigenvalues that give a tangible difference to the physics on $\bS^7$s with an exotic differential structure and hence give us a handle on how the physics can be different on topologically identical manifolds but with inequivalent differential structures.  

The general $G-$symmetric multi-instantons on symmetric spaces $G/H$ were studied by A. N. Schellekens \cite{Schellekens1,Schellekens2}. Here, we present an explicit construction in the case of the $4-$sphere $\bS^4\cong \SO(5)/\SO(4)$. We mostly follow the conventions and 
presentation of \cite{Dolan} with some precisions in the case of $\bS^4$. Using a decomposition $\mathfrak{so}(4)=\mathfrak su(2)\oplus \mathfrak su(2)$, each of the multi-instantons will be composed of a left $SU(2)-$multi-instanton and a right $SU(2)-$multi-instanton.
\subsection{Coset construction of $\bS^4$}
The $4-$sphere is seen as a coset space $\bS^4\cong \SO(5)/\SO(4)$.  {Then it is clear the symmetry of $\bS^4$ is $\SO(5)$.}  The 10 generators of $\SO(5)$ are labeled by $M,N,P,Q=1,\dots,10$ and the 6 generators of $\SO(4)$ are labeled by $a,b,c,d=1,\dots,6$, which of course is a closed subgroup of $\SO(5)$. Coordinate indices of the base manifold $\bS^4$ are labeled by  $\mu,\nu,\rho,\sigma=1,\dots,4$, and vierbein indices of $\bS^4$ are labeled by $m,n,p,q=1,\dots,4$.  The (anti-Hermitian) generators of $\SO(5)$ will be denoted by
\begin{align}
\{T_M\,,\,M=1,\dots,10\}
\end{align}
and its (totally anti-symmetric) structure constants $\{C^P_{MN}\quad|\quad\,P,M,N=1,\dots,10\}$ are defined by 
\begin{align}
    [T_M,T_N]=C^P_{MN} T_P\,.
\end{align}
 We fix a set of generators of 
$\so(4)$ as \begin{align}
\{T_a\,,\,a=1,\dots,6\}\,.
\end{align}
The remaining generators span the tangent space of $\bS^4$ at a fixed point,  $T(\SO(5)/\SO(4))$, and are denoted by
\begin{align*}
    \{T_\mu\,,\, \mu=1,\dots,4\}\,.
\end{align*}
Irreducible representations of $\SO(5)$ are labeled by two integers, $p,q$, $p\ge q\ge 0$,  with the corresponding representation noted as $(p,q)_5$.  Since $\SO(5)$ is compact, then in any representation $(p,q)_5$ there exist orthogonal generators $T_M((p,q)_5)$ satisfying
\begin{align}
    \Tr[T_M((p,q)_5)T_N((p,q)_5)]=-C_1^{\SO(5)}((p,q)_5)\,\delta_{MN}\,
\end{align}
$C_1^{\SO(5)}(R)$ is called the (second order) Dynkin index for the representation $(p,q)_5$.  It then follows, from the definition of the generators of $\SO(4)$, that $(p,q)_5$ induces a (possibly reducible) representation $R$ of $\SO(4)$, so that
\begin{align}
    \Tr[T_a(R)T_b(R)]=-C_1^{\SO(4)}(R)\,\delta_{ab}=-C_1^{\SO(5)}((p,q)_5))\,\delta_{ab},
\end{align}
which defines the normalizations of the generators of $\SO(5)$ and $\SO(4)$.  The quadratic Casimir operator in the representation $(p,q)_5$ is defined by
\begin{align}
 \sum_M T_M((p,q)_5) T_M((p,q)_5):=C_2^{\SO(5)}((p,q)_5)\, \bI.
\end{align}
It is related to $C_1^{\SO(5)}((p,q)_5)$ by 
\begin{align}
    C_2^{\SO(5)}((p,q)_5)=\frac{10}{\dim ((p,q)_5))} \,C_1^{\SO(5)}((p,q)_5)\,.
\end{align}
The quadratic Casimirs and the dimensions of the irreducible representations are well known and given respectively by
\begin{align}
    &C_2^{\SO(5)}((p,q)_5)=\frac{p^2+q^2}{2}+2p+q\\
    & \dim((p,q)_5)=\frac{1}{6}(p+q+3)(p-q+1)(p+2)(q+1)\,.
\end{align}

We have the following expression for the structure constants of $\SO(5)$, \cite{Schellekens2}:
\begin{align*}
    &C^a_{bc}\quad \text{ are the structure constants of $\SO(4)$}\\
    &C^\mu_{ab}=0\quad \text{ by closure of $\SO(4)$}\\
    &C^a_{\mu\nu}=\begin{cases}
        -\frac{1}{\sqrt{2}}\eta^{a}_{\mu\nu}\quad \text{if}\quad a=1,2,3\\
        -\frac{1}{\sqrt{2}}\overline{\eta}^{(a-3)}_{\mu\nu}\quad \text{if}\quad a=4,5,6
    \end{cases}\\
    &C^\mu_{\nu\gamma}=0\quad \text{since $\SO(5)/\SO(4)$ is a symmetric coset space}\,.
\end{align*}
where $\eta^{a}_{\mu\nu}$ and $\overline{\eta}^{(a-3)}_{\mu\nu}$ are the 't Hooft symbols, \cite{tHooft1,tHooft2}
\begin{align*}
&\eta_{\mu\nu}^{i}:=\epsilon_{\mu\nu i4} + \delta_{\mu i}\delta_{\nu 4}-\delta_{\mu 4}\delta_{\nu i},\quad &&\overline{\eta}_{\mu\nu}^{i}:=\epsilon_{\mu\nu i4} - (\delta_{\mu i}\delta_{\nu 4}-\delta_{\mu 4}\delta_{\nu i})\,
\end{align*}
which 'tHooft defined in his expression for the instanton gauge fields that are exact solutions of the Yang-Mills equations for the gauge group $\SU(2)$.  
\subsection{$\SO(5)$ invariant metric on $\bS^4$ }
For completeness, we record the $SO(5)$ invariant metric on $\SO(4)$.  On the $4-$sphere $\bS^4\cong\SO(5)/\SO(4)$, we put the standard $\SO(5)-$invariant Riemannian metric, the generators of $\SO(5)$ are the Killing vectors and the holonomy group is $\SO(4)$. The metric is obtained as follows. First, the $4-$sphere in $\R^5$ is defined by
 \begin{align*}
     \bS^4:=\big\{\,(z_1 , z_2 , \dots , z_5)\quad|\quad z_1^2+z_2^2+\dots+z_5^2=1\,\big\}\,.
 \end{align*}
 Consider the following local parametrization of $\bS^4$ in polar coordinates:
 \begin{align*}
     &z_1:=\sin\xi\sin\chi\sin\theta\cos\phi\\
     &z_2:=\sin\xi\sin\chi\sin\theta\sin\phi\\
     &z_3:=\sin\xi\sin \chi\cos\theta \qquad \qquad &\text{where}\qquad 0\leq \xi \leq \pi\,,\, 0\leq \chi \leq \pi\,,\, 0\leq \theta\leq \pi\,,\,0\leq \phi<2\pi\,.\\
     &z_4:=\sin \xi \cos \chi\\
     &z_5:=\cos \xi\,.
 \end{align*}
 The standard $\SO(5)-$invariant Riemannian metric on $\bS^4$ in these coordinates is
 \begin{align*}
     g_{\bS^4}&:=\d\xi \otimes \d\xi+\sin^2\xi\,(\, \d\chi\otimes\d\chi+\sin^2\chi\,\d\theta\otimes\d\theta+\sin^2\chi\sin^2\theta\,\d\phi\otimes\d\phi\,)\\
     &\equiv \sum_me^m\otimes e^m\,,
 \end{align*}
 where $\{e^m , m=1,\dots,4\}$ is the standard vierbein basis for this metric.
 The corresponding volume form is
 $$\d vol_{\bS^4}=\d z_1\wedge \d z_2\wedge\dots\wedge \d z_5= \sin^3\xi\sin^2\chi\sin\theta \,\d\xi\wedge\d\chi\wedge\d\theta\wedge\d\phi.$$
The spin connection of $\bS^4$ is defined by the equation $de^m+\omega^m_{\,\,\,\,   n}\wedge e^n=0$ and the curvature 2-form is defined by $R^m_{\,\, \,\, n}=d\omega^m_{\,\,\,\,   n}+\omega^m_{\,\,\,\,   p}\wedge \omega^p_{\,\,\,\,   n}$. In the standard vierbein basis $\{e^m , m=1, \dots, 4\}$, it is given by
   \beq
    \omega_{mn}=\frac{1}{1+z_5}(z_m \d z_n-z_n \d z_m)\label{7} \,.
\eeq
\subsection{Spherically symmetric gauge fields and the construction of spherically symmetric instantons on $\bS^4$}
We want to consider ``spherically'' symmetric connections on the bundles that define the exotic $\bS^7$s, $M^7_k$, simplified by the Kaluza-Klein reduction, because it is for just those gauge fields that we can use powerful group theoretical methods to address the spectrum of the Dirac operator.  %These are then $\so(4)-$connections, on a bundle corresponding to the manifolds defined by  $\SO(5)-$ invariant gauge potentials (the spherical symmetry), whose components are identified with those of the spin connections of $\bS^4$.  
Spherically symmetric solutions of the Yang-Mills equations (instantons) allow us to solve for the spectrum of the Dirac operator.  For general $h$ and $l$ there are no spherically symmetric instantons, \ie solutions of the Yang-Mills equations.    However, if one can find the appropriate embeddings, then the Dynkin indices of the embeddings \cite{Dynkin}, will be related to the topological invariants, $h,l$ of the connection and one can consider spherically symmetric instantons.  

A clear and simple example of this situation is given by Wilczek \cite{Wilczek}.  Here he considers a spherically symmetric instanton in an $\SU(3)$ gauge theory, but one that has topological charge 4.  The instanton resides in the $3\times 3$ spin 1 representation of {$\SO(3)$} embedded into $\SU(3)$.   The instanton can be continuously deformed into the upper left $\SU(2)$ sub block.  However, the spherical symmetry (and the fact that the configuration is a solution) is lost under the deformation.  One spatially separates the instanton into four charge 1 instantons residing in four different embeddings of the fundamental representation of $\SU(2)$ into $\SU(3)$, see \cite{Wilczek} for the details.   By local, topologically trivial $\SU(3)$ gauge transformations, each of these embeddings can then be gauge transformed into configurations corresponding to one specific embedding, say the standard embedding which corresponds to the $SU(2)$ subgroup of $\SU(3)$ sitting in the upper left $2\times 2$ block of the fundamental $3\times 3$ representation of $\SU(3)$.  Then the instantons can be brought together, giving rise to a charge 4 configuration in that standard embedding of $\SU(2)$ into $\SU(3)$.  This configuration does not satisfy the Yang-Mills equations, and it is not spherically symmetric.  There does exist a solution of the Yang-Mills equations for a charge 4 instanton in this  embedding of $\SU(2)$ in $\SU(3)$.    See for example the ADHM construction of the moduli space of all instanton solutions in the standard representation\cite{adhm}.   The upshot is that there do exist spherically symmetric instantons (\ie solutions of the Yang-Mills equations) for essentially any value of the instanton number, which is dependent on the Dynkin index of the embedded representation \cite{Dynkin}.  

These instantons reside in higher dimensional representations of $\SU(2)$ than the usual fundamental representation of $\SU(2)$, and can be thought of as embedded in a larger gauge group.  These higher gauge group instantons are bonafide instantons with higher instanton winding number, and are spherically symmetric.   We will use these spherically symmetric instantons to find the dependence of the spectrum of the Dirac operator on the manifolds with inequivalent differential structure. 

Correspondingly, we imagine we have a fundamental bundle of $\SO(4)$ instantons with charge $2h$ and $-2l$ in the left and right sector respectively.  These are not spherically symmetric in principle, however, if an appropriate representation of the gauge group is chosen, then depending on the embedding of the representation of $\SO(4)$ that we pick, we can get charge $2h$ or $-2l$ instantons with spherical symmetry.    We refer to \cite{Schellekens1} and \cite{Schellekens2} for more details. These embedded representations of $\so(4)$ will be denoted by $\cR_{h,l}$ which would not necessarily be an irreducible representation.   The irreducible representations of $\SO(4)$ are labelled by two half-integers, and $r,s$ with representation noted as $(r,s)_4$

We can now construct spherically symmetric $\SO(4)-$multi-instantons $\textcolor{black}{A}$ on $\bS^4$ with topological invariants $2h-2l$ (instanton number) and $h+l$ (Euler number) as follows. We consider the following $\so(4)-$valued singular $1-$form locally defined on $\bS^4$ :
 \begin{align*}
     A\equiv \sum_{r=1}^5A_r\,dz_r:=\sum_{m=1}^4\sum_{n=1}^4-\frac{1}{1+z_5
    }\eta^i_{mn}T^{[\textcolor{black}{h}]}_iz_n\,dz_m\,+\,\sum_{m=1}^4\sum_{n=1}^4-\frac{1}{1+z_5}\eta^i_{mn}T^{[\textcolor{black}{l}]}_iz_n\,dz_m\,,
 \end{align*}
 
 \noindent
where $\Big\{T^{[h]}_i\,,\,i=1,2,3\Big\}$ and $\Big\{T^{[l]}_i\,,\,i=1,2,3\Big\}$ are generators of the two $\su(2)$ factors in $\so(4)=\su(2)\oplus\su(2)$ which correspond to the  representations of $\SO(4)$ under which the fermions transform.  The left chirality spinors transform independently of the right chirality spinors, the corresponding gauge fields are self-dual and anti-self-dual, respectively.  We label the representations by $h$ and $l$, however, the representations of the left and right factors of $\SU(2)$ have the first Casimir (Dynkin index) given by $2h$ and $-2l$ respectively.  Additionally, the fermions carry intrinsic spin $\pm (1/2)$.  We take: 

\begin{align*}
[T^{[h]}_i,T^{[h]}_j]=\epsilon_{ijk}T^{[h]}_k\qquad,\qquad [T^{[l]}_i,T^{[l]}_j]=\epsilon_{ijk}T^{[l]}_k\qquad,\qquad [T^{[h]}_i,T^{[l]}_j]=0\,.
\end{align*}
By definition, they have the properties
\begin{align*}
\Tr\Big(T^{[h]}_iT^{[h]}_j\Big)=-\textcolor{black}{h}\,\delta_{i\,j}\qquad,\qquad\Tr\Big(T^{[l]}_iT^{[l]}_{j}\Big)=\textcolor{black}{l}\,\delta_{i\,j}\,,
\end{align*}

\noindent
 where we take $h>0$ and $l<0$ and which are the Dynkin indices of the embeddings of higher representations of $\so(4)=\su(2)\oplus\su(2)$ which determine $\cR_{h,l}$.
 
 For the specific case $h=2$, $l=-1$ for the $h=2$ we can take 
 \beq
 {T_i^{[ 2]}}_{jk}=-\epsilon_{ijk}\label{88}
 \eeq
 which satisfy
 \beq
 Tr\left(  T_i^{[ 2]} T_j^{[ 2]}\right)=-h\delta_{ij}=-2 \delta_{ij}
 \eeq
for the left component of $\so(4)$.  This representation of $\so(4)$ embeds smoothly with multiplicity 1 into the adjoint representation of $\so(5)$ noted as $(2,0)_5$ in standard notation.    For $l=-1$ we can take
 \beq
  T_i^{[ -1]}=-i\frac{\sigma^i}{2}\oplus -i\frac{\tau^i}{2}\label{99}
 \eeq
 where $\sigma^i$ and $\tau^i$ are independent Pauli matrices, which satisfy
 \beq
 Tr\left(  T_i^{[ -1]} T_j^{[ -1]}\right)=l\delta_{ij}= -\delta_{ij}
 \eeq
 for the right component of $\so(4)$.  This representation is unitarily equivalent to the right isoclinic factor of the fundmental representation of $\SO(4)$ that was described above, Eqn.\eqref{rightisoclinic}.   This representation embeds smoothly into the dimension 4 spinor representation of $\so(5)$ with multiplicity 1.   The multiplicity of the embedding of any representation of $\SO(N-1)$ in an irreducible representation of $\SO(N)$ is well known to be exactly 1, \cite{multiplicity}.  The manifold with $h=2$, $l=-1$ satisfies $h+l=1$ but $h-l=3\ne \pm 1\,\,{\rm modulo}\,\,7$ and hence describes an exotic sphere.
 
\subsection{Spectrum of the Dirac operator (squared)}

In this section we will compute the spectrum of the squared Dirac operator on $\bS^4$ in the gauge fields that we have constructed for all values of $h$ and $l$.   {The spectrum does contain a contribution that is an irrelevant constant, however large that it might be.  This contribution comes from the lowest energy mode that lives on the fibre.  This contribution to the spectrum, is large, it behaves like $1/R$ where $R$ is the radius of the fibre, however, it is a constant as far as the exotic differential structures are concerned.  It does not change from one exotic sphere to the next, therefore, it is irrelevant for the present analysis.  The spectrum on the full $\bS^7$ depends on which exotic sphere one starts with, and constrains the mass/energy spectrum for fermions on $\bS^4$ after Kaluza-Klein reduction. } The fibre being $\bS^3$, it is known that it admits no harmonic spinors (zero modes).  However, the contribution to the spectrum of the Dirac operator from its action in the fibre is independent of the gauge fields that is, how the $\bS^3$ is twisted as one moves over the base, $\bS^4$.  The contribution to the eigenvalues will behave like $1/R$ or even a higher power of $1/R$, but it will not depend on the gauge field and hence on the coordinates of the base.  To the coordinates on $\bS^3$ the gauge field is just a constant, and can be removed by a gauge transformation, and does not affect the spectrum.  Thus the part of the Dirac operator that depends on the coordinates of the fibre just gives a universal constant contribution, universal in the sense that it does not depend on which exotic sphere we are considering.  It does not depend on the differential structure of the $\bS^7$, \ie on the gauge fields that are effectively coming from the non-trivial differential structure.  It is actually possible to add other monopole type gauge fields into the fibre to obtain exact zero modes of the part of the Dirac operator on the fibre, however this is besides the point.   The contribution to the spectrum of the Dirac operator coming from the small $\bS^3$ is universal in the sense that it does not depend on the differential structure.    This contribution does not contribute to the changes in the spectrum  due to inequivalent differential structure which is the part of the spectrum that we want to understand.  

We consider the standard Riemannian metric on $\bS^4$. After Kaluza-Klein reduction, the Einstein-Yang-Mills-Dirac action on the compactified space-time $\bS^4$ is given by : 
\begin{align*}
        \mathcal{S}_{\cE-\mathcal{YM}-\cD}=\int_{\bS^4} \Big(R_{\bS^4}+R_{\bS^3}+\frac{1}{2}\mathscr{L}_{\cY\cM}[A]+\overline{\psi}(i\cD_A)\psi\Big)\,\d vol_{\bS^4}\,.
    \end{align*}
The Dirac operator on $\bS^4$ in a $\SO(4)-$gauge field background $A=A_r dz_r$ is given (using our conventions for the indices) by 
\begin{align}\label{eq:Relativistic Dirac-Landau operator onS4}
    \mathcal{D}_A&=\gamma^l e^r_l\Big(\partial_r+\frac{1}{4}\omega_{mn,r}\gamma^{mn}+iA_r\Big)\,.
\end{align}
Here $\{e^m\equiv e^m_r dz_r\,,\,\,m=1,\dots,4\}$ form the standard orthonormal coframe for $\bS^4$ and the components of the spin connection $1-$form of $\bS^4$ are given by Eqn.\eqref{7},
\begin{align*}
    \omega_{mn}=\frac{1}{1+z_5}(z_m \d z_n-z_n \d z_m)\,
\end{align*}
and $\gamma^{mn}:=\frac{1}{2}[\gamma^m,\gamma^n]$, with  $\gamma^m$ the  usual Dirac gamma matrices satisfying  $\{\gamma^m,\gamma^n\}=2\delta^{mn}$.  Then, the Dirac equation for $\psi$ is
\begin{align}\label{eq:Dirac-equation}
    i\gamma^l e^r_l\Big(\partial_r+\frac{1}{4}\omega_{mn,r}\gamma^{mn}+iA_r\Big)\psi=0\, .
\end{align}
We will aim to find the spectrum of the Dirac operator $\cD_A$ (squared).   

Exploiting the assumed spherical symmetry of the gauge field,  Dolan \cite{Dolan} has found general formulas for the spectrum of the square of the Dirac operator on a homogeneous space.  The square of an eigenvalue $\lambda$
\beq
i\cal D_A\psi=\lambda \psi
\eeq
of the Dirac operator  will of course be an eigenvalue, $\lambda^2$,  of the square of the Dirac operator $(i\mathcal{D}_A)^2$, however, the converse, that $\pm\sqrt{\lambda^2}$ will correspond to eigenvalues of the Dirac operator, does not necessarily follow.  

Dolan's results are obtained as follows.  We note that his work, as he himself points out, leans heavily on previous work of Salam-Strathdee \cite{SalamStrathdee} and was well understood in the mathematics literature \cite{kobayashi-nomizu}.  Recording the more general case, let $G/H$ be a Riemannian homogeneous coset space, with $G$ and $H$ compact Lie groups and $G$ simple, such that its isometry group is $G$ and its holonomy group is $H$. Let $t_M , M=1\dots,\dim G$ be the anti-hermitean generators of $G$, with $[t_M , t_N]=C_{MN}^P t_P$, and $t_a , a=1 , \dots , \dim H$ will denote the generators of  $H$. Let $A$ be a $G-$symmetric gauge potential on $G/H$ and (using our conventions for the indices)
    $$\mathcal{D}_A:=\gamma^\alpha e^{\mu}_\alpha\Big(\partial_\mu+\frac{1}{4}\omega_{\delta\beta,\mu}\gamma^{\delta\beta}+i A_\mu\Big)$$ is the Dirac operator on $G/H$, where $\{e^\alpha_\mu dx^\mu\,,\,\alpha=1,\dots,\dim (G/H)\}$ form an orthonormal coframe for $G/H$. Here $\alpha,\beta=1,\dots, \dim G/H$ are orthonormal indices and $\mu , \nu=1,\dots,\dim G/H$ are coordinate indices.   The orthonormal 1-forms can be taken as the Maurer-Cartan 1-forms on the whole of $G$
\beq
g^{-1}d g=e^At_A
\eeq
such that
\beq
de^A=\frac{1}{2}C^A_{\,\,\,BC}e^B\wedge e^C.
\eeq
The set of 1-forms separate into a subset $e^\alpha$ for a $G-$invariant metric on $G/H$ and the remaining $e^a$ can be expanded as $e^a=\Pi^a_{\,\,\,\alpha} e^\alpha$ on the manifold $G/H$.   The ensuing spin connection is obtained from
\beq
de^\alpha +\omega^\alpha_{\,\,\,\beta}\wedge e^\beta =0
\eeq
yielding the curvature 2-form
\beq
R^\alpha_{\,\,\,\,\beta}=\frac{1}{2}R^\alpha_{\,\,\,\,\beta\gamma\delta}e^\gamma\wedge e^\delta=\frac{1}{2}C^\alpha_{\,\,\,\,\beta a}C^a_{\,\,\,\,\gamma\delta}e^\gamma\wedge e^\delta .
\eeq
We can calculate $(i\mathcal{D}_A)^2$ to find
\beq
(i\mathcal{D}_A)^2= -D_\alpha D^\alpha +\frac{R}{4}\mathbb I +\frac{i}{2}F_{\alpha\beta}\gamma^{\alpha\beta}\label{77}
\eeq
where $R$ is the Ricci scalar and $D_\alpha D^\alpha=-\Delta$ is the $G$ symmetric Dirac Lapacian acting on spinors including the spin connection and the gauge connection defined on $G/H$.   For the specific, spherically symmetric gauge fields, all three terms on the RHS of Eqn.\eqref{77} are mutually commuting and therefore can be simultaneously diagonalized.   One can compute and find
\beq
[D_\alpha,D_\beta ]=i F^a_{\alpha\beta}t_a+\frac{1}{4}R_{\alpha\beta\gamma\delta}\gamma^{\gamma\delta}
\eeq
where $t_a$ are the generators of the chosen representation of $H$.   

The notion of spherical symmetry means that we choose a metric and connection that are $G$ invariant.    In our case, $G=\SO(5)$ and $H=\SO(4)$ giving $G/H=\SO(5)/\SO(4)=\bS^4$ as the base and the fibre $H=\SO(4)\simeq \SU(2\times \SU(2)/\mathbb Z_2$ is 6-dimensional.  The gauge field being spherically symmetric means that a Killing vector $K$, generates via the Lie derivative just a gauge transform, $F$ is invariant up to a gauge transformation
\beq
{\cal L}_K F=g^{-1}F g.
\eeq
Such an invariance is obtained by taking the gauge connection to be equal to the spin-connection, which is possible as the gauge group is the holonomy group $H$.  The gauge field strength is given by
\beq
F^a=\frac{1}{2}F^a_{\,\,\,\,\alpha\beta}e^\alpha\wedge e^\beta=\frac{1}{2}C^a_{\,\,\,\,\alpha\beta}e^\alpha\wedge e^\beta .
\eeq
The Riemann tensor is covariantly conserved hence so is the field strength
\beq
D_\alpha F^a_{\,\,\,\,\beta\gamma}=0
\eeq
and with this choice for the gauge field, it is easy to verify
\beq
[D_\alpha ,D_\beta ]=C^a_{\alpha\beta}\left(  \mathbb I\otimes t_a - \frac{1}{4}C_{a\gamma\delta} \gamma^{\gamma\delta}\otimes\mathbb I \right) .
\eeq
However, interestingly, $T_a= - \frac{1}{4}C_{a\gamma\delta} \gamma^{\gamma\delta}$ gives a representation of the holonomy gauge group $H$
\beq
[T_a,T_b]=C^c_{\,\,\,\, ab}T_c
\eeq
which then implies the commutator
\beq
[ D_\alpha ,D_\beta ] =C^a_{\,\,\,\alpha\beta} D_a
\eeq
where 
\beq
D_a=\mathbb I\otimes t_a + T_a\otimes\mathbb I .
\eeq
Then we can write the Dirac Laplacian as
\beq
\Delta=-D_\alpha D_\alpha = -D_AD_A+D_aD_a
\eeq
but these are just the quadratic Casimirs of $G$ and $H$ respectively.  These Casimirs simply depend on the representation of the groups that is being considered.  Therefore we can write
\beq
\Delta=C_2(G,\cdot ) -C_2(H,D_a) .
\eeq
where the $C_2(G,\cdot)$ indicates any representation of $G$ that contains the representation $D_a$ of $H$.  As we scan over all such representations, we get all the possible eigenvalues of the Dirac Laplacian.  This is completely analogous to the action of the spherical Laplacian on the spherical harmonics, the result there is $l(l+1)$ for the eigenvalue of the spherical Laplacian, depending on which spherical harmonic is considered.  The eigenvalue is obtained from pure group theory, there is actually no necessity to solve for the eigenfunctions of the partial differential operator given by the Laplacian!  Therefore, in total we have
        \begin{align*}
    (i\mathcal{D}_A)^2=C_2^G-C_2^H+\frac{1}{8}R_{G/H}\,.
\end{align*}

In our case, we consider a symmetric homogeneous space $\SO(5)/\SO(4)\cong\bS^4$ of unit radius (endowed with its standard $\SO(5)-$invariant Riemannian metric) with holonomy group $\SO(4)$ the scalar curvature is
 $$R_{G/H}=R_{\bS^4}=12\,$$ 
giving a contribution of $\frac{3}{2}$ as a cosmological constant.  The irreducible representations $(p,q)_5$ of $\so(5)$ have quadratic Casimirs (eigenvalues)
 $$C_2^{\SO(5)}((p,q)_5)=\frac{p^2+q^2}{2}+2p+q\,.$$
 Hence, the full spectrum of the squared Dirac operator $\mathcal{D}_A^2$ on $\bS^4$ in any of our symmetric gauge field backgrounds constructed before will have the form
\begin{align*}
    E^{[h,l]}_{p,q}={\frac{p^2+q^2}{2}+2p+q-C_2^{\SO(4)}(\cR_{h,l})+\frac{3}{2}}\,,
\end{align*}
where the quadratic Casimir operator $C_2^{\SO(4)}(\cR_{h,l})$ also denotes its eigenvalues in the representation $\cR_{h,l}$. Here, there is the constraint that $p\geq q$ and that the irreducible representations $(p,q)_5$ of $\so(5)$ used to compute the spectrum  should contain the (embedded) representation $\cR_{h,l}$ of $\so(4)$.  Additionally, the total eigenvalue will have independent contributions from the left and right sectors.  

It was shown by Yang \cite{Yang}, in his prescient study of $\SU(2)$ monopoles on $\bS^4$, that the representations of $\SO(5)$ which contain a given representation $I$ of the $\SU(2)$, satisfy 
\beq
p-q=2 I
\eeq
where $I$ is the total ``isospin'' of the fermion, comprising of the combination of the gauge ``isospin'' $J$ and the intrinsic ``isospin'' of the fermion, $1/2$, \cite{Yang}.   Thus $I=J\pm (1/2)$, $p=q +2I$ and $C_2^{\SO(4)}(\cR_{J,0})=J(J+1)$.    It is also well known that the multiplicity of the embedding is 1, whenever the embedding is possible.  This gives for the case of Eqn.\eqref{88}
\beq
    E^{[J,0]}_{q+2I,q}={q^2}+q(2I+3) +2I^2 +4I -J(J+1)
\eeq
There will be an independent contribution for the left-handed spinors and the right-handed spinors, transforming according to representation labeled by $h$ and $l$ respectively.  The relationship between $J$ and $h$ or $l$ can be slightly complicated.  In our example, $h=2$ corresponds to an irreducible representation $(1,0)_4$ of  $\SO(4)$ while for $l=-1$, the representation corresponds to the reducible representation $(0,1/2)_4\oplus (0,1/2)_4$.  Thus the complete eigenvalue will have a representation in $(p,q)_5$ of $\SO(5)$ for the left-handed sector in which is embedded the representation labelled by $h$ of $\SO(4)$ and a representation in $(p',q')_5$ of $\SO(5)$ for the right-handed sector in which is embedded the representation labelled by $l$ of $\SO(4)$.  Thus the full spectrum of eigenvalues will be
\beq
\lambda^2(q,I,q',I')= E^{[h,0]}_{q+2I,q}+ E^{[0,l]}_{q'+2I',q'}+3/2
\eeq

\underline{Example} 1 : If we start the Kaluza-Klein reduction process  with the standard sphere $\bS^7\cong M^7_{1}$, where $h=1$ and $l=0$, \ie $h+l=1$ and $k=h-l=1$.   Then we find that the spectrum for the Dirac operator will be in the representation $h=1$, whch corresponds to $(\frac{1}{2},0)_4=\frac{1}{2}$ of $SU(2)_L$ and the representation for $l=0$, which corresponds to $(0,0)_4$ or the trivial representation of $\SU2)_R$.  The left and right handed spinors then will be independently appended by a representation of $\SO(5)$ which permits the embedding of  the given representation of $\SO(4)$.  As the right handed spinor is trivial, we will have simply $I'=\pm (1/2)$ and then we get
\bea
E^{[1,0]}_{q+2I,q}+E^{[0,0]}_{q'+2I',q'}+3/2={q^2}+q(2((1/2)\pm (1/2))+3) +2((1/2)\pm (1/2))^2 +4((1/2)\pm (1/2)) -(1/2)((1/2)+1)\nonumber\\
+{q'^2}+q'(2(\pm (1/2))+3) +2(\pm (1/2))^2 +4(\pm (1/2)) -(0)(0+1) +3/2\nonumber\\
{~}
\eea
In this case, the reduced/effective 4D theory is just the standard Einstein-Yang-Mills theory on $\bS^4$ with $\SU(2)$ Yang-Mills gauge group and our $\SO(4)-$multi-instanton reduces to the BPST $1-$instanton. The Milnor's bundle is just the standard quaternionic Hopf fibration.\\

\underline{Example} 2 : If we start the Kaluza-Klein process with an exotic $7-$sphere $M^7_{3}$, where $h=2$ and $l=-1$, \ie $h+l=1$ and $k=h-l=3$.   The extra term corresponding to the eigenvalues of $C_2^{\SO(4)}(\cR_{2,-1})$ will depend on the integers $h=2$ and $l=-1$.  Clearly the spectrum will not be the same as for the theory on the standard sphere.  In this case, the  isopin for the left-handed sector will have $J=1$ so that $I=1\pm (1/2)$ while for the right handed sector, the isospin of the direct sum of two spin one-half representations will act in concert and be $J=1/2$ giving $I'=(1/2)\pm (1/2)$.  Then the eigenvalues of the Dirac operator (squared) will be
\bea
E^{[2,0]}_{q+2I,q}+ E^{[0,-1]}_{q'+2I',q'}+3/2={q^2}+q(2(1\pm (1/2))+3) +2(1\pm (1/2))^2 +4(1\pm (1/2)) -1(1+1)\nonumber\\
+{q'^2}+q'(2((1/2) \pm (1/2))+3) +2((1/2) \pm (1/2))^2 +4((1/2)\pm (1/2)) -(1/2)((1/2)+1)+3/2.
\eea

\section{Conclusions and Future Work}   
 {Thus, we see directly how different choices of smooth structures on the 7-sphere affect the energy/mass spectrum for fermions. The spectrum is altered due to global topological reasons. Diffeomorphically inequivalent smooth structures on topological 7-spheres exist because maps between these manifolds, although continuous, cannot always be made differentiable everywhere for certain values of the parameters $h$ and $l$. The failure of differentiability occurs at least at one point \cite{Milnor}. Our results demonstrate that the spectrum of the Dirac operator on an exotic $\bS^7$ differs from that on the standard $\bS^7$.}
 {
These results may have broader applications in quantum mechanics, condensed matter physics, and Kaluza-Klein supergravity. In condensed matter systems, the ground state degeneracies of higher-dimensional quantum Hall effects are linked to the Atiyah-Singer index theorem for spinors in background gauge fields. Observable consequences of inequivalent differential structures could lead to previously unforeseen physical phenomena. Since effective theories in condensed matter often mirror higher-dimensional physics, these findings could prove especially relevant.}
 {
Another possible application is in the study of moduli spaces arising in quantum information. For example, the moduli space of the two-qubit states is topologically $\bS^7$ \cite{rmrd}, as the normalized general state
$$
|\psi\rangle = \alpha |\!\uparrow\uparrow\rangle + \beta |\!\uparrow\downarrow\rangle + \gamma |\!\downarrow\uparrow\rangle + \delta |\!\downarrow\downarrow\rangle
$$
satisfies the constraint $|\alpha|^2 + |\beta|^2 + |\gamma|^2 + |\delta|^2 = 1$, leading to a 7-dimensional parameter space. While this space is topologically a sphere, its differential structure could, in principle, be exotic. A natural question is what connection should be defined over this space, and whether it corresponds to a nontrivial Pontryagin class imposing and exotic differential structure. We speculate\footnote{in preparation} that the Berry connection may play this role and could potentially reflect an exotic smooth structure on $\bS^7$.
}

\section{ACKNOWLEDGEMENTS}
We thank NSERC, Canada for financial support and Richard MacKenzie, Ben Webster(Perimeter) and Benedict Williams (UBC)  for useful discussions.  We thank Richard Easther and the University of Auckland, Department of Physics, Auckland, New Zealand, for hospitality where this work was finally completed and written up.
 
\bibliographystyle{apsrev}
\bibliography{ref}

\end{document}